\documentclass[a4paper,11pt]{article}
\pdfoutput=1 

\usepackage{jcappub} 

\usepackage[T1]{fontenc} 

\usepackage{lineno}

\newcommand{\ICfilesloc}{\href{http://icecube.wisc.edu/science/data/IC79_solarWIMP_data_release}{http://icecube.wisc.edu/science/data/IC79\_solarWIMP\_data\_release}}

\title{\boldmath Constraining Secluded Dark Matter models with the public data from the 79-string IceCube search for dark matter in the Sun}

\author[a,1]{M. Ardid, \note{Corresponding author.}}
\author[a]{I. Felis,}
\author[b]{A. Herrero,}
\author[a]{and J.A. Mart\'{\i}nez-Mora}

\affiliation[a]{\scriptsize{Institut d'Investigaci\'o per a la Gesti\'o Integrada de les Zones Costaneres (IGIC) - Universitat Polit\`ecnica de Val\`encia. C/  Paranimf 1 , 46730 Gandia, Spain.}}

\affiliation[b]{\scriptsize{Institut de Matem\`atica Multidisciplinar - Universitat Polit\`ecnica de Val\`encia. Cam\'{\i} de Vera s/n , 46022 Val\`encia, Spain.}}

\emailAdd{mardid@fis.upv.es}
\emailAdd{ivfeen@upv.es}
\emailAdd{aherrero@mat.upv.es}
\emailAdd{jmmora@fis.upv.es}

\abstract{The 79-string IceCube search for dark matter in the Sun public data is used to test Secluded Dark Matter models. No significant excess over background is observed and constraints on the parameters of the models are derived. Moreover, the search is also used to constrain the dark photon model in the region of the parameter space with dark photon masses  between 0.22 and $\sim$ 1 GeV and a kinetic mixing parameter $\varepsilon \sim 10^{-9}$, which remains unconstrained. These are the first constraints of dark photons from neutrino telescopes. It is expected that neutrino telescopes will be efficient tools to test dark photons by means of different searches in the Sun, Earth and Galactic Center, which could complement constraints from direct detection, accelerators, astrophysics and indirect detection with other messengers, such as gamma rays or antiparticles.}

\begin{document}
\maketitle
\flushbottom


\section{Introduction}
\label{sec:intro}

There is strong cosmological and astrophysical evidence for the existence of Dark Matter (DM) in the Universe. The observations indicate that DM, about 26\% of the total mass-energy of the Universe, is non-baryonic, non-relativistic and not subject to electromagnetic interactions \cite{01}. In the framework of the Weakly Interacting Massive Particles (WIMPs) paradigm, the visible baryonic part of a galaxy is embedded in the DM halo. In the most common scenario, WIMPs can scatter elastically with matter and become trapped in massive astrophysical objects such as the Sun\cite{02}. There, DM particles could self-annihilate, reaching equilibrium between capture and annihilation rates over the age of the Solar System. The standard scenario assumes that the products of DM annihilation are Standard Model (SM) particles, which interact with the interior of the Sun and are largely absorbed \cite{03}. However, during this process, high-energy neutrinos \footnote{In the paper neutrino will mean neutrino plus anti-neutrino, unless explicitly stated otherwise.} may be produced, which can escape and be observed by neutrino detectors, such as IceCube or ANTARES. Limits on WIMP DM annihilation in the Sun have been reported already in neutrino telescopes: ANTARES \cite{1,AntDm}, Baksan \cite{2}, Super-Kamiokande \cite{3} and IceCube \cite{4,IC79}. 

An alternative hypothesis is based on the idea that DM may live in a dark sector with its own forces, so it is ``secluded'' from SM particles and that the annihilation is only possible through a metastable mediator ($\phi$), which subsequently decays into SM states \cite{5,6,7,8,9}. For example, the dark matter's stability may be ensured not by some discrete parity imposed by hand,
but simply by being the lightest fermion in the dark sector.  If
the dark sector contains an Abelian gauge symmetry, dark
electromagnetism, the dark photon and the Standard Model (SM) photon
will generically mix kinetically.  This mixing is of special
interest because it is one of the few ways for a dark sector to
interact with the known particles through a renormalizable interaction
and it is non-decoupling: a particle charged under both dark and
standard electromagnetism induces this interaction at loop-level, and
the effect is not suppressed for very heavy particles.  In this way,
this is a simple example for dark matter models with light
mediators \cite{DP-Earth}. This kind of models retains the thermal relic WIMP DM scenario while at the same time  explain the positron-electron ratio observed by PAMELA \cite{10}, FERMI \cite{11}, and AMS-II \cite{12}. In the Secluded Dark Matter (SDM) scenario, the presence of a mediator dramatically changes the annihilation signature of DM captured in the Sun. If the mediators live long enough to escape the Sun before decaying, they can produce fluxes of charged particles, $\gamma$-rays or neutrinos \cite{13,14} that could reach the Earth and be detected. In many of the secluded dark matter models, $\phi$ can decay into leptons near the Earth. The signature of leptons arising from $\phi$ decays may differ substantially from other DM models. Assuming that the DM mass ($\sim$ 1 TeV) is much greater than the $\phi$ mass ($\sim$ 1 GeV) the leptons are boosted. If these leptons are muons, which is the less constrained case to explain the positron-electron ratio \cite{15,16,17}, the signature in the vicinity of the detector would be two almost parallel muon tracks. In Ref. \cite{18} this possibility is discussed and the expected sensitivity for the IceCube neutrino telescope is calculated. 
The use of neutrino telescopes has also been proposed to detect the decay of dark photons coming from DM annihilation in the Center of the Earth \cite{DP-Earth}. The same authors also proposed to study dark photons coming from DM annihilation in the Center of the Sun by means of positron detection with AMS \cite{DP-Sun}.
Moreover, even for short-lived mediators that decay before reaching the Earth, neutrinos from the products of mediator decays could be detected in neutrino telescopes as well. Other authors also explored the possibility that mediators may decay directly into neutrinos \cite{20}. In this case, the neutrino signal could be enhanced significantly compared to the standard scenario even for quite short-lived mediators since the mediators will be able to escape the dense core of the Sun, where high-energy neutrinos can interact with nuclei and be absorbed. The fact that the solar density decreases exponentially with radius facilitates neutrinos injected by mediators at larger radii to propagate out of the Sun. Recently, an indirect search for SDM using the 2007-2012 data recorded by the ANTARES neutrino telescope was reported \cite{ANTSDM} setting constraints on the dark matter mass and the lifetime of the mediator for this kind of models. 
Text from \cite{ANTSDM} have been used under CC BY 3.0, adapted to this paper.

In this paper, a similar analysis is done using the public available data from the 79-string IceCube search for dark matter in the Sun. We look for neutrinos from decays of dimuons or dipions produced by mediators that decay before reaching the Earth, or neutrinos produced by mediators that decay directly to neutrinos. Afterwards, this search is also used to constraint the dark photon model. 
Finally we show that the detection of neutrinos coming from the Sun is also a good approach for testing the dark photon model.  


\section{A search for SDM in the Sun with the IceCube 79-string configuration public data}

\subsection{The IceCube 79-string configuration detector and the data sample}
\label{icdescription}

The IceCube neutrino observatory~\cite{IC_review} is a neutrino telescope situated at the South Pole, installed in the ice at depths of between 1450\,m and 2450\,m, instrumenting a total volume of one cubic kilometre. Digital Optical Modules (DOMs) arranged on vertical strings record the Cherenkov light induced by relativistic charged particles, including those created by neutrinos interacting with the ice. The detection of photons by DOMs with good timing synchronization allows the reconstruction of the directions and energies of the secondaries.
In its 79-string configuration, 73 strings have a horizontal spacing of 125\,m and a vertical spacing of 17\,m between DOMs. The six remaining strings are located near the central string of IceCube and feature a reduced vertical spacing between DOMs of 7\,m and higher quantum efficiency photomultiplier tubes.  These strings along with the seven
surrounding regular strings form the DeepCore subarray. The horizontal distance between strings in DeepCore is
less than 75\,m. 

In the analysis described in this paper, the public available data from a search for WIMP dark matter annihilation in the Sun with the IceCube 79-string configuration~\cite{IC79} is used, particularly the data corresponding to the 136-day sample (`winter high-energy' event selection, WH), which  has no particular track-containment requirement and aims to select upward-going muon tracks. 
Full event data from the analysis of Ref. \cite{IC79} can be found at \ICfilesloc, including angles (the reconstructed zenith and azimuth, and the angle of muon candidate event relative to Sun position, $\Psi$), the energy proxy, $N^{\rm c}_\mathrm{chan}$, values, the paraboloid sigmas (uncertainty on the reconstructed direction) and the event time stamp.  Effective areas and volumes, along with $N^{\rm c}_\mathrm{chan}$ and angular responses, can also be found at the same location.

The background consists of muons arising in single or coincident air showers as well as atmospheric neutrinos. They can be estimated by scrambling real data at the final analysis level. The IceCube Collaboration also released the estimation of the background distributions, that is, the total number of background events per event selection with the background probability density functions (PDFs) of the angular distribution between the reconstructed track and the position of the Sun in 1-degree bins. The PDFs of the background in $N^{\rm c}_\mathrm{chan}$ was also provided by IceCube Collaboration.

\subsection{Signal estimation and optimisation of the event selection criteria}\label{sec:signal}

      
To evaluate the IC79 sensitivity to the neutrinos that are the final decay products that arrives at Earth, the effective areas for neutrinos as a function of the $N^{\rm c}_\mathrm{chan}$  and $\Psi$ have been used. For this, it is necessary to know the energy spectra of neutrinos arriving at the detector. In case of mediators decaying into muons, the final spectra have been obtained from Michel's spectra of neutrinos from muon decay and taking into account the boost \cite{25}. The assumption that after oscillations all neutrino flavours arrive at Earth with the same proportion has been made. This aspect results in a factor of 2/3 for the muon neutrinos with respect to the parent muons since both a muon neutrino and an electron neutrino result from a muon decay and these neutrinos are spread to all flavours. A similar process is used for mediators decaying into pions, considering the neutrinos produced by the decay of the pion and the daughter muon. For mediators decaying directly to neutrinos and assuming long mediator lifetimes with respect to the time required to exit from the Sun's core, the neutrino spectra are almost flat in the energy range under study \cite{20}. Here, the idea of all flavours being present with the same ratio is based not only on oscillations but also on equal production.


In order to avoid any bias in the event selection, sensitivity studies for the different SDM models for different DM masses as a function of the observables (angular separation of the track with respect to Sun's direction, $\Psi$ and the energy proxy, $N^{\rm c}_\mathrm{chan}$) were done. Since there was not a significant gain in the energy proxy parameter for the DM mass range used in this study 0.1 -- 10 TeV and it was not our aim to fine-tune for the different scenarios it was decided not to cut on the energy proxy parameter. Finally, the selected cut value for $\Psi$ was the one which gave the best sensitivity for the neutrino flux using the Model Rejection Factor (MRF) method \cite{26}. 
It consists of finding the angle cut which provides, on average, the best flux upper limit taking into account the observed background and the efficiency for a possible signal flux. The procedure to obtain the sensitivity for a neutrino flux coming from DM annihilations in the Sun is described in Ref. \cite{1}. 
For each DM mass and annihilation channel, the chosen $\rm \Psi_{cut}$ is the one that minimise 
the average 90\% confidence level (CL) upper limit on the muon neutrino flux 
flux, $\rm \overline{\Phi}_{\nu_{\mu}}$, 
defined as

\begin{equation}
\rm{\overline{\Phi}_{\nu_{\mu}} = \frac{\bar{\mu}^{90\%}}{\bar{A}_{eff}(M_{WIMP}) \times T}} \, ,
\label{mrfeq}
\end{equation}

\noindent where  $\rm \bar{\mu}^{90\%}$ is the average upper limit of the background at 
$90$\% CL computed using a Poisson distribution in the Feldman-Cousins approach~\cite{27}  and $\rm T$ is the total lifetime for each detector configuration. The effective area averaged  over the neutrino energy, $\rm{\bar{A}_{eff}(M_{\rm WIMP})}$, is defined as:

\begin{equation}
\rm{\bar{A}_{eff}(M_{\rm WIMP}) =
\sum_{\nu,\bar{\nu}} \left ( \frac{\displaystyle\int_{E_{\nu}^{th}}^{M_{\rm WIMP}} A_{eff}(E_{\nu,\bar{\nu}}) \, \frac{dN_{\nu,\bar{\nu}}}{dE_{\nu,\bar{\nu}}} dE_{\nu,\bar{\nu}}}
{\displaystyle\int_{0}^{M_{\rm WIMP}}\frac{dN_{\nu}}{dE_{\nu}} dE_{\nu} \,+\, \frac{dN_{\bar{\nu}}}{dE_{\bar{\nu}}} dE_{\bar{\nu}}} \right )}  \, ,
\label{aeffeq}
\end{equation}

\noindent where $\rm E_{\nu}^{th}=43$~GeV is the energy threshold for neutrino detection in IC79, $\rm M_{WIMP}$ 
is the WIMP mass, $\rm dN_{\nu,\bar{\nu}}/dE_{\nu,\bar{\nu}}$ is the energy spectrum of the (anti-)neutrinos, and $\rm A_{eff}(E_{\nu,\bar{\nu}})$, which is provided by IceCube, is the effective area of  IC79 as a function of the (anti-)neutrino energy for tracks coming from the direction of the Sun.

\begin{figure}[tbp]
\begin{center}
\includegraphics[width=12cm, angle=270]{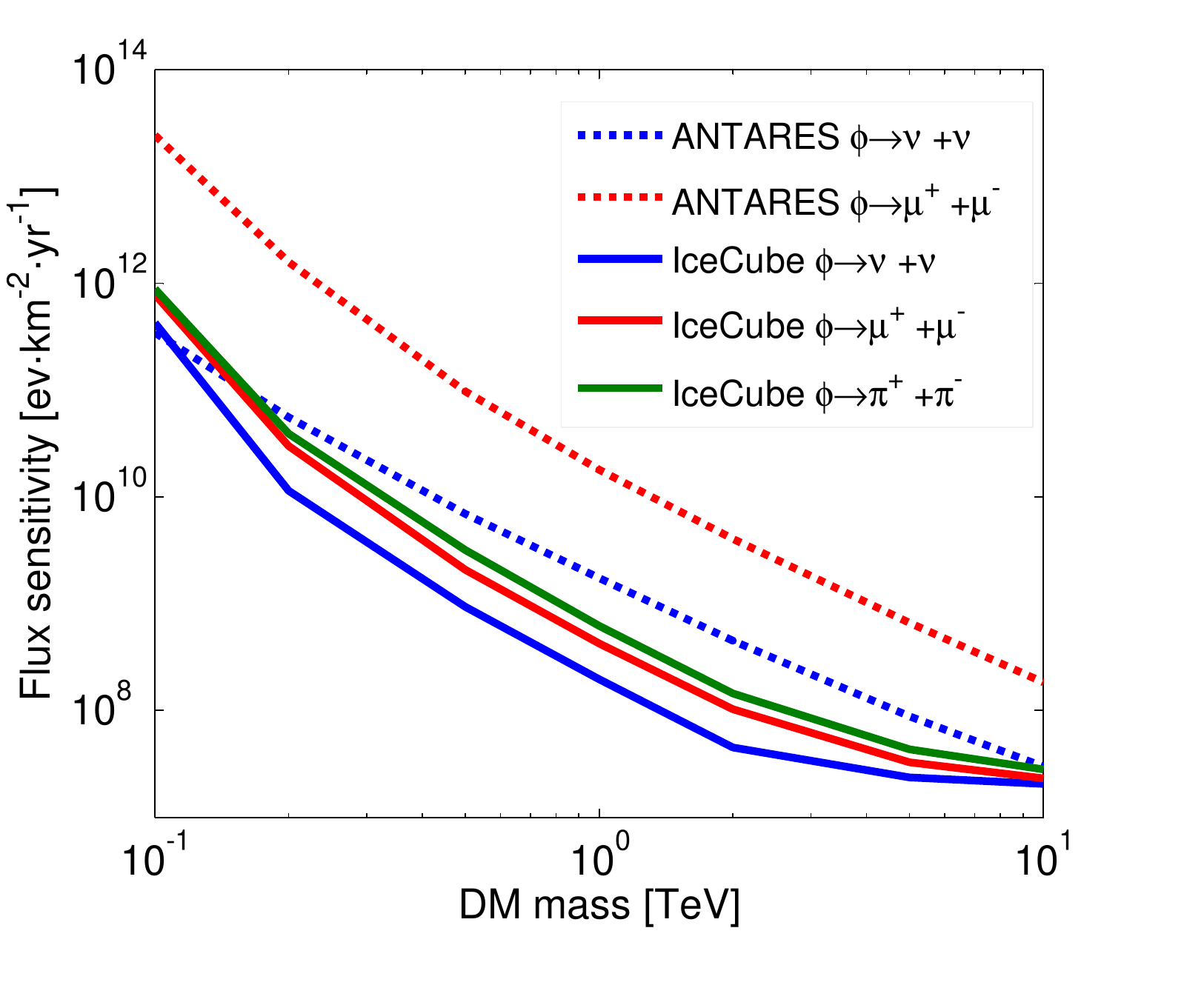}
\end{center}
\caption{\label{fig:1} Sensitivities to muon neutrino flux for DM annihilation in the Sun for possible SDM channels from this search with IC79 compared with the ANTARES search \cite{ANTSDM}.}
\end{figure}

The $\Psi_{\rm cut}$ values, the expected background events, $\rm \bar{\mu}^{90\%}$ and the sensitivities for some DM masses and channels are presented in Table \ref{tab:1}. The observed events within $\Psi_{\rm cut}$ is also shown. The global flux sensitivities obtained as a function of the DM mass for different SDM channels are shown in Figure \ref{fig:1}. They are compared with the ones obtained from an ANTARES search \cite{ANTSDM}.

\subsection{Results and discussion} \label{sec:disc}

After the optimisation of the flux sensitivities and selection of $\Psi_{\rm cut}$ using the MRF, we proceed to observe the number of neutrino events in the data that come from the Sun direction. The number of events in data and the expected ones from background matched within the statistical fluctuations expected, see Table \ref{tab:1}. Therefore, no significant excess was observed, and 90\% Confidence Level (CL) upper limit values were used to constrain the model using the Feldman-Cousins approach \cite{27}. 

\begin{table}[tbp]
\centering
\begin{tabular}{ccccccc}
\hline
DM mass (TeV) & $\phi$ decay channel & $\Psi_{\rm cut}$(degree) & $N_{\rm back}$ & $\bar{\mu}^{90\%}$ & $N_{\rm obs}$ & $\mu_{90\%}$ UL\\
\hline
0.1 & $\mu, \pi, \nu$ & 3.0 & 60.4 & 12.8 & 70 & 25\\
0.2 & $\mu, \pi, \nu$ & 1.5 & 15.2 & 8.0 & 19 & 12.4\\
0.5 & $\mu, \pi, \nu$ & 1.2 & 9.8 & 6.7 & 13 & 10.3\\
1 & $\mu, \pi$ & 1.1 & 8.2 & 6.3 & 10 & 8.3\\
1 & $\nu$ & 1.0 & 6.8 & 5.8 & 8 & 7.2\\
2 & $\mu, \pi$ & 0.9 & 5.6 & 5.4 & 6 & 5.9\\
2 & $\nu$ & 0.8 & 4.5 & 5.0 & 3 & 3.2\\
5 & $\mu, \pi, \nu$ & 0.8 & 4.5 & 5.0 & 3 & 3.2\\
10 & $\mu, \pi, \nu$ & 0.8 & 4.5 & 5.0 & 3 & 3.2\\
\hline
\end{tabular}
\caption{\label{tab:1} $\Psi_{\rm cut}$ used for the different DM masses and $\phi$ decay channel. The expected background $N_{\rm back}$, $\bar{\mu}^{90\%}$, the number of observed events $N_{\rm obs}$ and the resulting 90\% confidence level upper limit are also shown ($\mu_{90\%}$ UL).}
\end{table}

Similarly to Eq. \ref{mrfeq}, upper limits on the neutrino fluxes for the different channels and DM masses were derived using the upper limits, $\mu_{90\%}$, obtained in Table \ref{tab:1} and the corresponding effective areas. 
Following the arguments given in Ref. \cite{18,25}, the neutrino flux at Earth can be translated into DM annihilation rate in the Sun. Assuming a 100\% branching ratio for the $\phi\rightarrow\mu^+ +\mu^-$ (or $\phi\rightarrow\pi^+ +\pi^-$) decay channel of the mediator, and taking into account the solid angle factor, the decay probabilities and the ratio considering flavour oscillations, the following relationships between the annihilation rate, $\Gamma$, and the muon neutrino flux, $\Phi_{\nu_\mu}$, are obtained: 
\begin{itemize}
\item[] Case $\phi\rightarrow\mu^+ +\mu^-$
\begin{equation}\label{gamma1}
\Gamma=\frac{4\pi D^2\Phi_{\nu_\mu}}{\frac{8}{3}(e^{-R_{Sun}/L}-e^{-D/L})},
\end{equation}
\item[] Case $\phi\rightarrow\pi^+ +\pi^-$
\begin{equation}\label{gamma2}
\Gamma=\frac{4\pi D^2\Phi_{\nu_\mu}}{\frac{12}{3}(e^{-R_{Sun}/L}-e^{-D/L})},
\end{equation}
\end{itemize}
where $D$ is the distance between the Sun and the Earth; $R_{Sun}$ is the radius of the Sun; and $L$ is the mediator's decay length, $L=\gamma c \tau$, i.e. the product of the mediator's lifetime, $\tau$, the speed of light, $c$, and the relativistic boost factor $\gamma= m_\chi/m_\phi$.

For the case in which mediators decay directly into neutrinos, if the lifetime of the mediator is small, the final energy spectrum of neutrinos would be quite similar to the case of typical DM searches \cite{20}. Therefore, only the situation in which the mediator lifetime is long enough has been considered, so that the absorption of neutrinos in the Sun becomes negligible ($L>10^{5}$ km). In this scenario, the relationship between $\Gamma$ and $\Phi_{\nu_\mu}$ is \cite{25}:
\begin{itemize}
\item[]Case $\phi\rightarrow\nu +\nu$
\begin{equation}\label{gamma3}
\Gamma=\frac{4\pi D^2\Phi_{\nu_\mu}}{\frac{4}{3}(1-e^{-D/L})}.
\end{equation}
\end{itemize}

\begin{figure}[tbp]
\centering  
\includegraphics[width=6.25cm, angle=270]{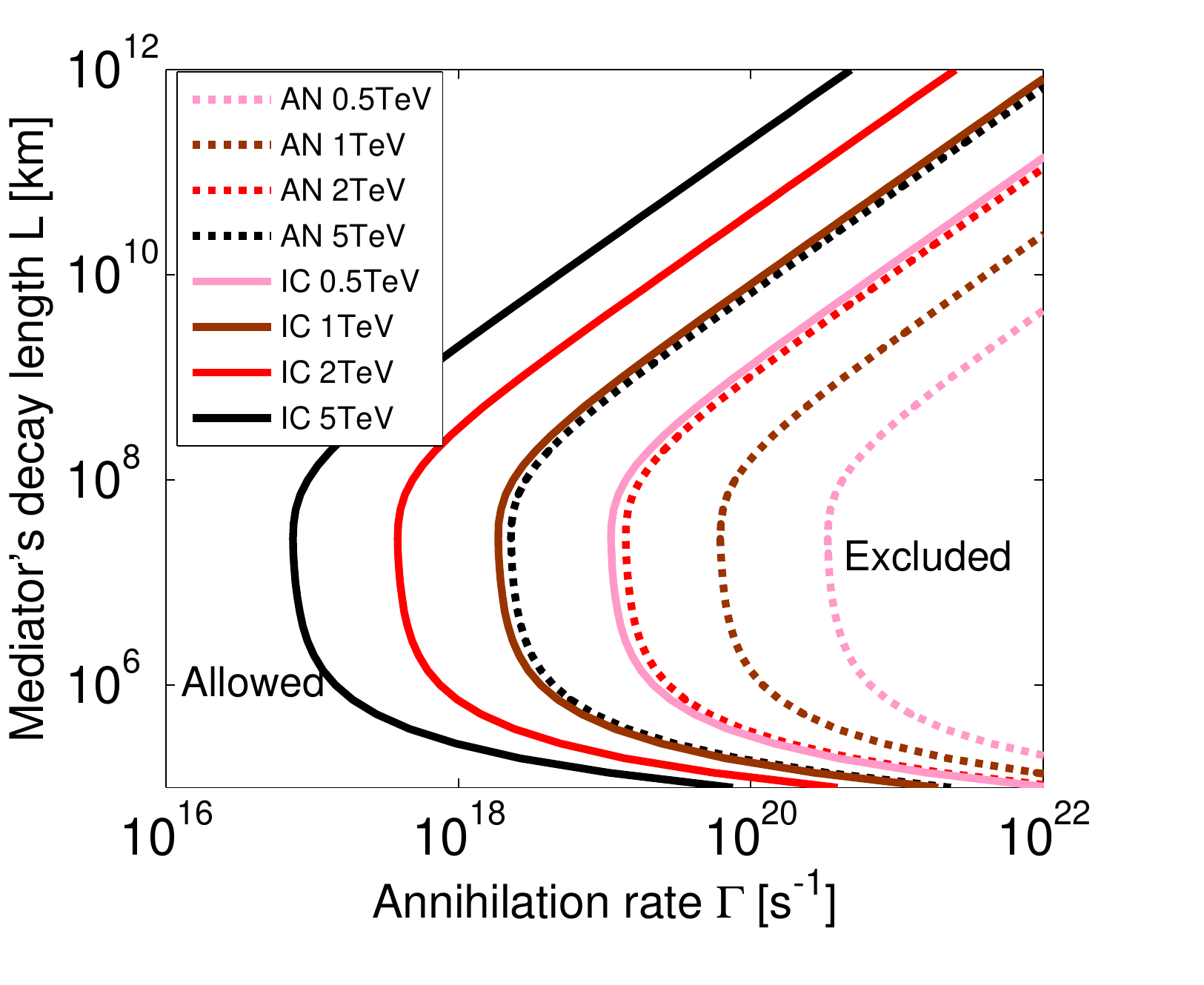}
\hfill
\includegraphics[width=6.25cm,angle=270]{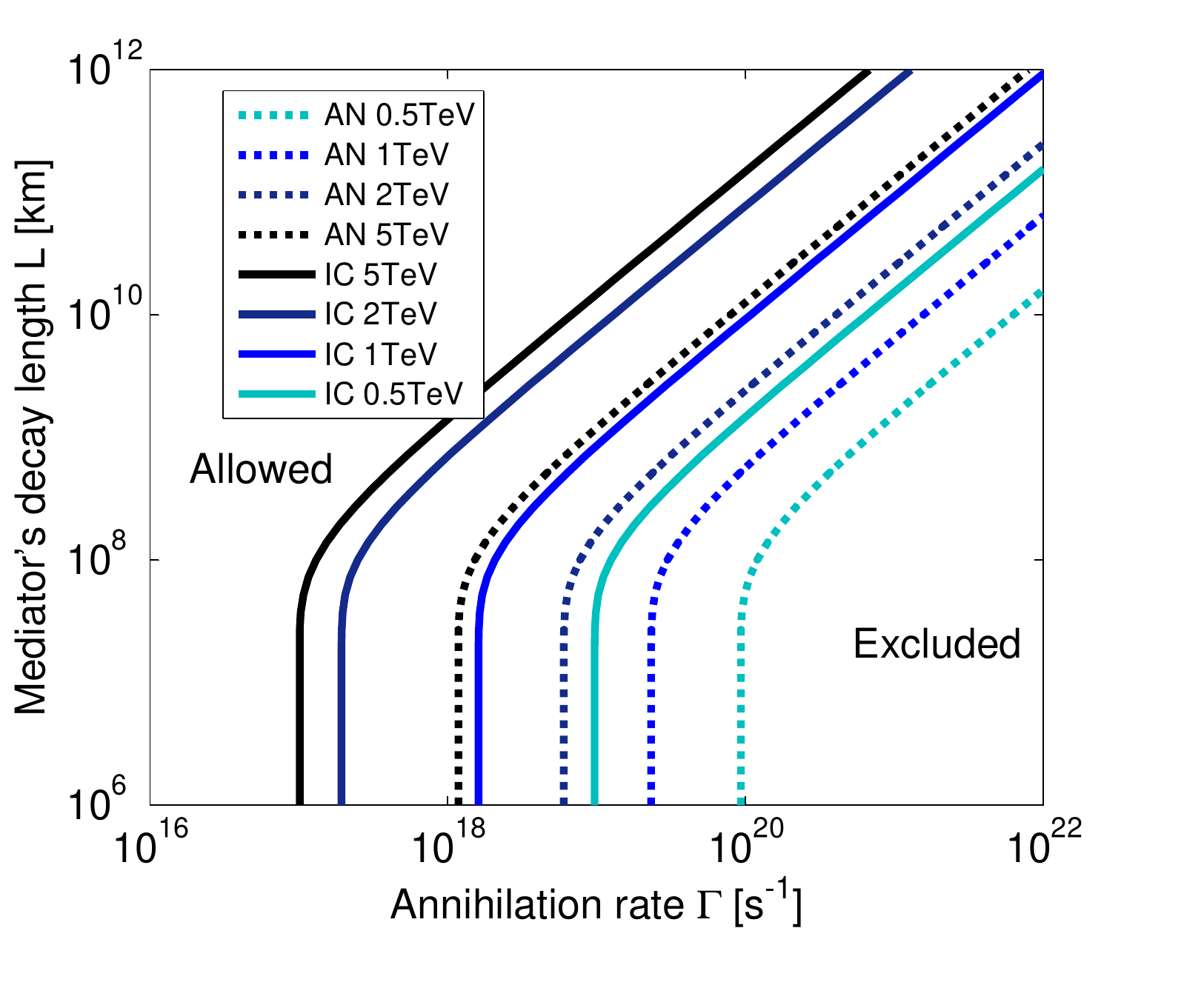}
\caption{\label{fig:3} IC79 exclusion limits for the SDM cases studied by products of DM annihilation in the Sun through mediators decaying into muons (left) and into neutrinos (right) as a function of the annihilation rate ($\Gamma$) and the decay length ($L = \gamma c\tau$). ANTARES results from \cite{ANTSDM} are also shown. The right side of the line corresponding to a certain DM mass is the excluded region.}
\end{figure}

Constraints on the annihilation rates as a function of mediator lifetime and dark matter mass have been obtained. For example, Figure \ref{fig:3} shows the IC79 exclusion limits for the mediator decay into muons (left) and into neutrinos (right).  The limits are compared to a previous search from ANTARES \cite{ANTSDM}.

\begin{figure}[tbp]
\centering  
\includegraphics[width=8cm,angle=270]{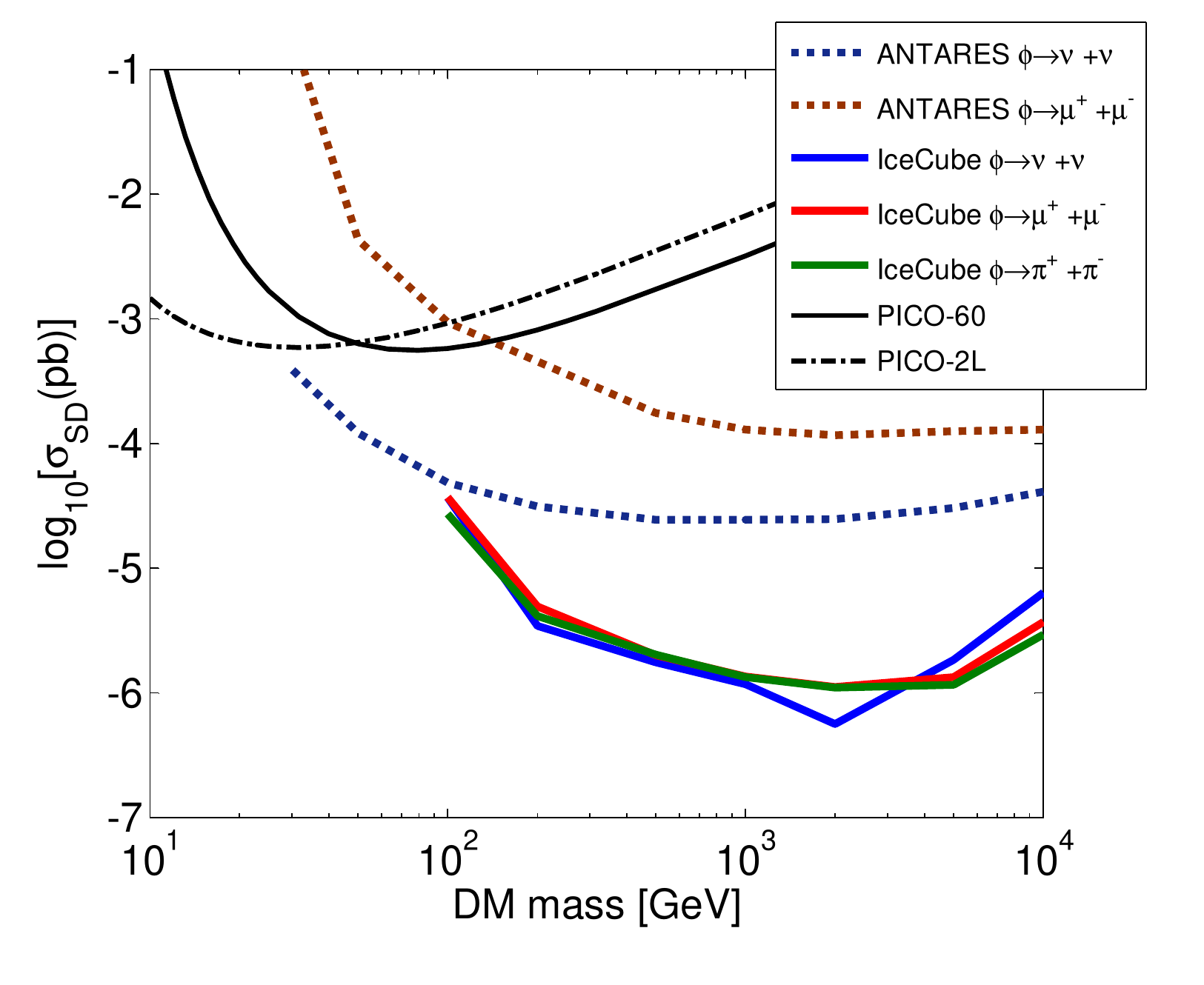}
\includegraphics[width=8cm,angle=270]{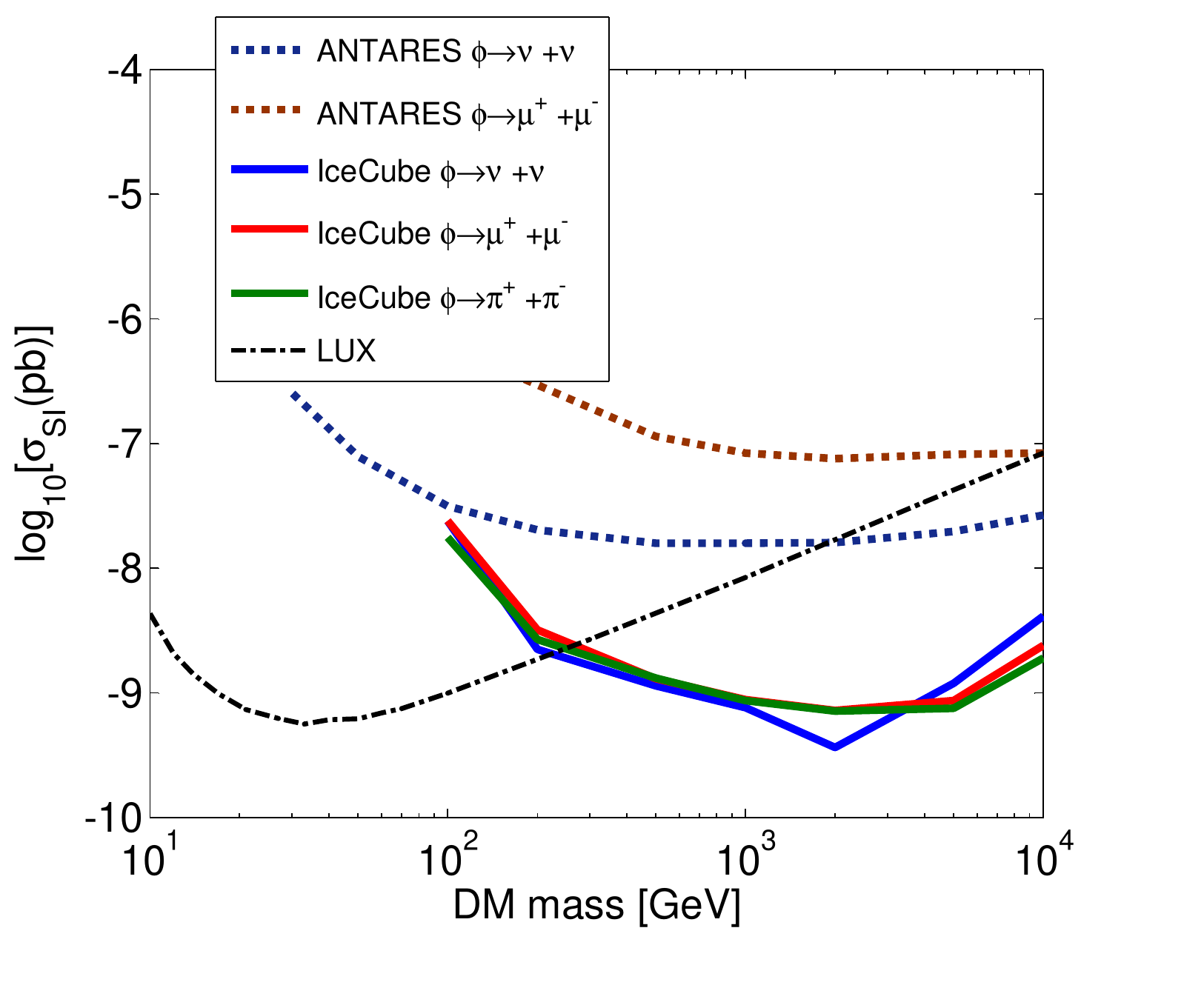}
\caption{\label{fig:4} IC79 90\% CL upper limits on WIMP-proton cross section as a function of WIMP mass. The top panel refers to spin-dependent and the bottom one to spin-independent WIMP interactions. A decay length of $2.8\times 10^7$ km is considered. The results from an ANTARES search \cite{ANTSDM}, as well as the current bounds from PICO \cite{30,30b,30c} and LUX \cite{LUX} are also shown.}
\end{figure}

Limits on the DM-nucleon interaction can also be derived for these cases. Taking the usual assumption that there is equilibrium of the DM population in the Sun, i.e., the annihilation balances the DM capture, $\Gamma =C_{DM}/2$ , according to \cite{28} the capture is approximately
\begin{equation}\label{cdm}
C_{DM}=10^{20} {\rm s}^{-1} \left(\frac{1 {\rm TeV}}{m_\chi}\right)^2  \frac{2.77\sigma_{SD}+4270\sigma_{SI}}{10^{-40} {\rm cm}^2} , 
\end{equation}
where, $\sigma_{SD}$ and $\sigma_{SI}$ are the spin-dependent (SD) and spin-independent (SI) cross sections, respectively. The limits on the SD and SI WIMP-proton scattering cross-sections are derived for the case in which one of them is dominant. The sensitivity in terms of annihilation rates depends on the lifetime of the mediator $\phi$. To assess the potential to constrain these models, lifetime values for which the sensitivities are optimal have been assumed. The lifetime of the mediator for the best sensitivity has to be long enough to ensure that the mediator escapes the Sun, but not so long that it decays before reaching the Earth. Figure \ref{fig:4} shows the IC79 nucleon--WIMP cross section limits derived for the SDM scenario with the products of DM annihilation in the Sun through mediators decaying into di-muons (red), di-pions (green) and directly into neutrinos (blue) for a mediator's decay length of $2.8 \times 10^7$ km. This is the optimal value according to Eq.~\ref{gamma1} and \ref{gamma2}. The limit  dependence with decay length $L$ is small for $R_{Sun}<< L << D$, but can be very large outside this region. The results from an ANTARES search \cite{ANTSDM}, as well as the current bounds from direct detection are also shown in Figure \ref{fig:4}. Thus, for sufficiently long-lived but unstable mediators, the limits imposed to these models are much more restrictive than those derived in direct detection searches for the case of spin-dependent interaction. In the case of spin-independent interactions, the direct detection search is more competitive for low and intermediate masses, but the SDM search becomes more competitive for larger masses ($> 0.2$ TeV). 

\section{Constraints for dark photon model}\label{DarkPhoton}

\subsection{Dark Matter Interactions Through a Dark Photon}

In this SDM model the mediator is the dark photon $A'$ which is the gauge boson of a broken U(1) symmetry that kinetically mixes with the hypercharge boson. When the dark photon mass is light, the mixing with the $Z$ boson is negligible and this system may be treated as a mixing between the photon and the dark photon. There are four independent parameters of the theory in the simplest model of dark matter, $\chi$, interacting through dark photons $A'$:  the masses of dark matter, $m_\chi$, and dark photon, $m_{A'}$,  the kinetic mixing parameter, $\varepsilon$, and $\alpha_X = g_X^2/(4\pi)$, being $g_X$ the dark U(1) gauge coupling. Typically, the $\alpha_X$ is also fixed by requiring $\chi$ to saturate the observed dark matter density through thermal freeze out, so $\alpha_X =\alpha_X^{\text{th}} \simeq 0.035 (m_\chi / \rm{TeV})$.
Alternatively, the maximum allowed coupling is set by bounds on distortions to the cosmic microwave background, and then $\alpha_X^\text{max} \simeq 0.17 (m_X / \rm{TeV})^{1.61}$ for the mass range considered \cite{DP-Earth} .
With a choice of $\alpha_X$, the model
is completely determined by the first 3 parameters 
and the dark photon decay length is not an independent parameter. In the limit where $m_\chi \gg m_{A'} \gg m_e$,  it is
\begin{align}
	L
	&= 
	R_{Sun} \, \text{Br}(A'\to e^+e^-)
	\left( \frac{1.1 \times 10^{-9}}{\varepsilon} \right)^2
	\left( \frac{m_\chi/m_{A'}}{1000} \right)
	\left( \frac{100\text{ MeV}}{m_{A'}} \right),
	\label{eq:decay:length}
\end{align}
where $R_{Sun} = 7.0 \times 10^{10}\text{ cm} = 4.6 \times 10^{-3}\text{ au}$ is the radius of the Sun, and $B_e= \rm{Br}(A'\to e^+e^-)$ the branching ratio to $e^+e^-$. For $m_{A'} < 2
m_{\mu}$, $B_e = 100\%$.  As $m_{A'}$ increases above $2 m_{\mu}$, the
$A' \to \mu^+ \mu^-$ decay mode opens up rapidly, 
and $B_e$ drops to 50\% at $m_{A'} \sim$ 300~MeV.  For 500~MeV $< m_{A'} <$ 1~GeV,
$B_e$ and $B_{\mu}$ are nearly identical and typically vary between
15\% and 40\%, with the rest made up by decays to hadrons, mainly pions~\cite{Buschmann:2015}.

\subsection{Exclusion regions using IceCube 79-string data}

The limits derived previously for neutrino fluxes in SDM can be translated into constraints to Dark Photon model. Parameter regions which predict a larger flux of neutrinos than the flux limit obtained will be excluded. Two conditions have to be accomplished for this. Firstly, there should be sufficient  capture and annihilation rate of DM in the Sun for producing the neutrino flux, and thus the nucleon-WIMP cross section should be large enough. To evaluate this aspect the approximate expression for short range regime has been used \cite{nonWIMP}:

\begin{equation}\label{eq:sigmaDD_dark}
\sigma_D^{\rm SI}    \approx 
10^{-40} \text{ cm}^2
\left( \frac{\epsilon}{10^{-4}} \right)^2      
\left( \frac{g_{_X}}{0.1} \right)^2     
\left( \frac{1 \text{ GeV}}{m_\chi} \right)^4 .
\end{equation}

Secondly, the branching ratio of $A'$ decay into muons (or pions) and its decay length according to Eq. \ref{eq:decay:length} should be adequate to produce a large  neutrino flux. 
 
\begin{figure}[tbp]
\centering  
\includegraphics[width=9cm, angle=270]{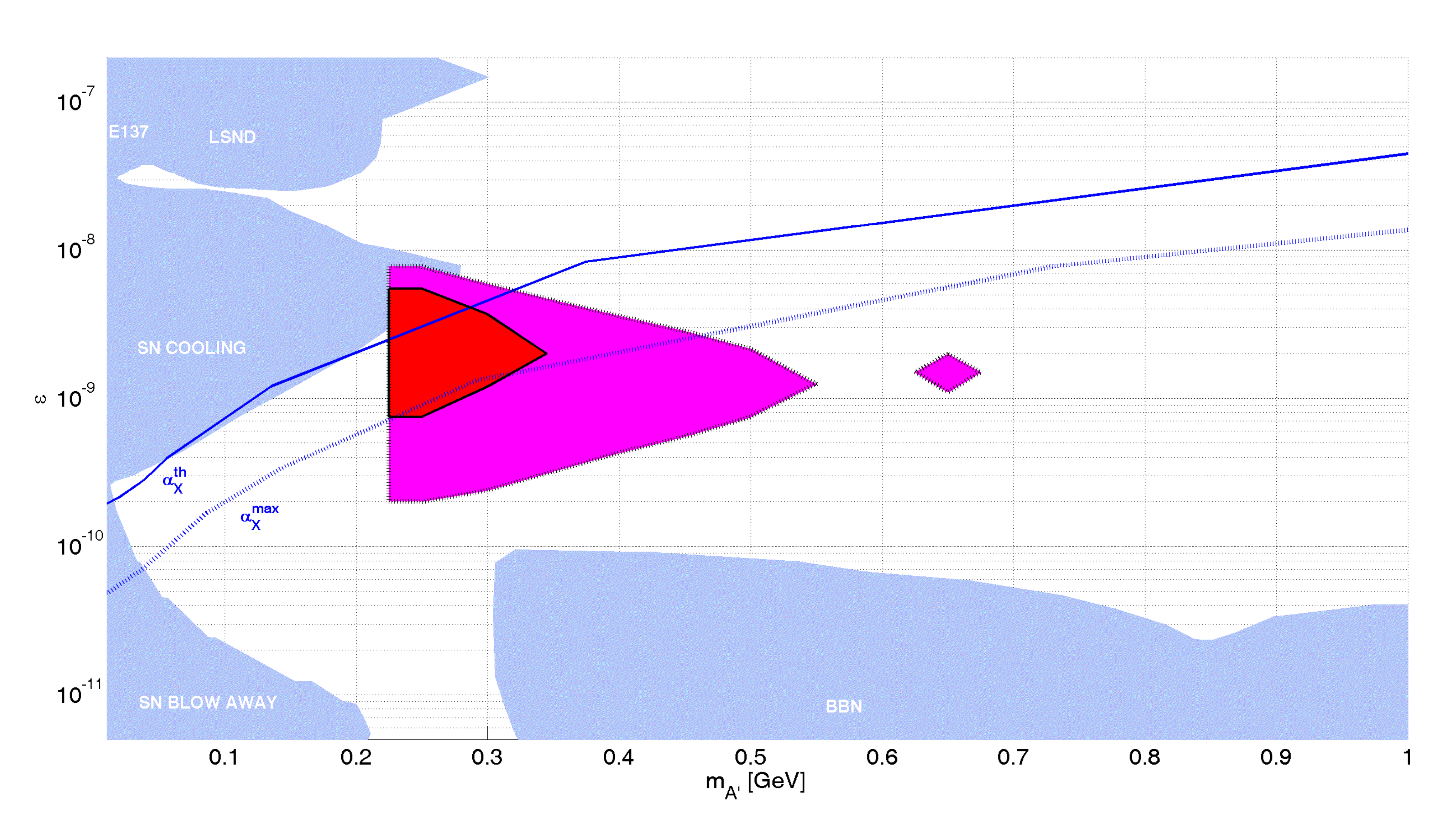}
\caption{\label{fig:DP1} Excluded regions in the $(m_{A'}, \varepsilon)$ plane by this analysis (using IC79 data) for a DM mass of 10 TeV and $\alpha_X = \alpha_X^\text{th}$ (red) or $\alpha_X^{\text{max}}$ (magenta). Bounds from direct detection and regions probed by accelerators and astrophysics are also shown, taken from \cite{DP-Sun}.
}
\end{figure}

\begin{figure}[tbp]
\centering  
\includegraphics[width=9cm, angle=270]{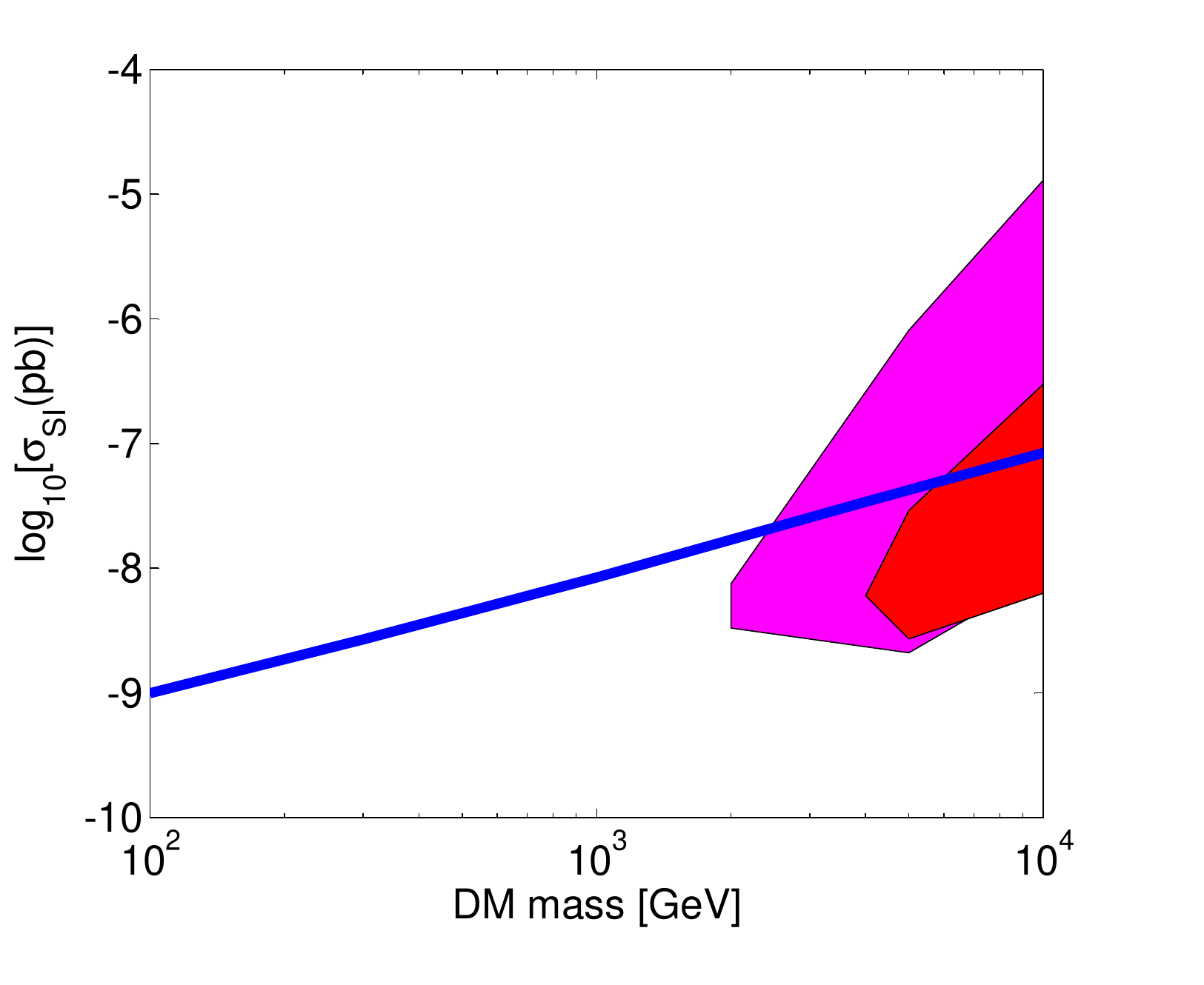}
\caption{\label{fig:DP2} Constraints in the $(m_X, \sigma)$ plane from this analysis (IC79 data)for $m_{A'} = 250$~MeV and assuming $\alpha_X = \alpha_X^\text{th}$ (red) or $\alpha_X^{\text{max}}$ (magenta). The bounds from LUX \cite{LUX} are also shown.
}
\end{figure}

In Figure \ref{fig:DP1} the excluded regions using the IC79 data are presented in the $(m_{A'}, \varepsilon)$ plane considering a DM mass of 10 TeV and compared to other bounds from direct detection, accelerator or astrophysic results. Here $\alpha_X = \alpha_X^\text{th}$ is considered for the red region and $\alpha_X^{\text{max}}$  for the magenta region. Naturally,  the test presented is only possible for  $m_{A'} > 2 m_{\mu}$. Most of the regions are constrained by the decay into muons since the energy spectrum of the neutrinos produced is hard and the branching ratio for $m_{A'}\sim 250$~MeV is quite large. However, the little magenta region on the right, $m_{A'}\sim 650$~MeV, is constrained through the $A'$ decay into pions since the branching ratio for this channel is large and the energy spectrum of the neutrinos produced is not so hard as in the case of muons, but hard enough. As it can be observed from the figure, there are some unconstrained regions that can be constrained using existing or future neutrino telescopes \cite{KM3-LOI,NewIc}. 
For smaller DM masses the parameter region constrained is also smaller, in fact, this analysis does not constrain DM masses below 1 TeV. This can be better observed by looking at Figure \ref{fig:DP2}, where the constraints of this analysis are presented in the $(m_X, \sigma)$ plane for a $m_{A'} = 250$~MeV considering $\alpha_X = \alpha_X^\text{th}$ (red region) or $\alpha_X^{\text{max}}$ (magenta region). Last bounds from LUX \cite{LUX} are also presented for comparison.

Recently, Cirelli et al. \cite{Cirelli:2016} have released a paper studying this kind of models and seeking the different constraints from direct detection, accelerators, astrophysics and indirect detection through gamma rays and antiprotons, which also constrain part of the region unconstrained. Thus, indirect detection can be a good tool to constrain the model and a larger region can be excluded
by combining different messengers (gammas, antiprotons and neutrinos) in a complementary and more reliable way since the astrophysical assumptions and uncertainties can be different for different searches.

\section{Conclusions}\label{concl}

The 79-string IceCube search for dark matter in the Sun public data has been analysed to test Secluded Dark Matter models. No significant excess of neutrinos over background is observed and constraints on the parameters of the models are derived. For long-lived but unstable mediators, the limits derived are more restrictive than those from direct detection. Moreover, the search has also been used to constrain the dark photon model in the region of the parameter space with dark photon masses between 0.22 and $\sim$ 1 GeV and a kinetic mixing parameter $\varepsilon \sim 10^{-9}$, which remains unconstrained. These are the first constraints of dark photons from neutrino telescopes, showing that neutrino telescopes can be efficient tools to test dark photons by means of different searches in the Sun, Earth and Galactic Center that could complement constrains from direct detection, accelerators, astrophysics and indirect detection with other messengers, such as gamma rays or antiparticles.

\acknowledgments

We acknowledge the financial support of Plan Estatal de Investigaci\'on, ref. FPA2015-65150-C3-2-P (MINECO/FEDER), Consolider MultiDark CSD2009-00064 (MINECO) and of the Generalitat Valenciana, Grant PrometeoII/2014/079. We would like to thank the colleagues J.D. Zornoza, C. Rott, J.L. Feng, J. Smolinsky and P. Tanedo for the fruitful discussions and comments about this work.

\end{document}